\documentclass[article,twocolumn,pra,amsmath,amssymb,superscriptaddress,longbibliography]{revtex4-2}
\usepackage[english]{babel}
\usepackage{amssymb}
\usepackage{dcolumn}
\usepackage{bm}
\usepackage{amsmath}
\usepackage{tabularx}
\usepackage{graphicx}        
\usepackage{makecell}                             
\graphicspath{{pict/}{fig2_2D_asymmetric}}

\usepackage{xcolor}
\usepackage{multirow}
\usepackage{dcolumn}
\usepackage{bm}
\usepackage{float}

\usepackage[normalem]{ulem}
\usepackage{comment}
\usepackage{epstopdf}
\usepackage{hhline}

\usepackage[raggedleft]{titlesec}
\titleformat{\section}{\bf\large}{\thesection.\,}{0.24em}{}
\titlespacing{\section}{0cm}{0.5cm}{0cm}
\titleformat{\subsection}{\bf\large}{\thesubsection.\,}{0.24em}{}
\titlespacing{\subsection}{0cm}{0.5cm}{0cm}

\begin{document}

\onecolumngrid

\title{Activating non-Hermitian skin modes by parity-time symmetry breaking}
\author{Zhoutao Lei}
\affiliation{Guangdong Provincial Key Laboratory of Quantum Metrology and Sensing $\&$ School of Physics and Astronomy, Sun Yat-Sen University (Zhuhai Campus), Zhuhai 519082, China}
\author{Ching Hua Lee}\email{phylch@nus.edu.sg}
\affiliation{Department of Physics, National University of Singapore, Singapore 117551, Republic of Singapore}
\author{Linhu Li}\email{lilh56@mail.sysu.edu.cn}
\affiliation{Guangdong Provincial Key Laboratory of Quantum Metrology and Sensing $\&$ School of Physics and Astronomy, Sun Yat-Sen University (Zhuhai Campus), Zhuhai 519082, China}

\begin{abstract}\noindent{
\textbf{Abstract:}
Parity-time ($\mathcal{PT}$) symmetry is a cornerstone of non-Hermitian physics as it ensures real energies for stable experimental realization of non-Hermitian phenomena. In this work, we propose $\mathcal{PT}$ symmetry as a paradigm for designing rich families of higher-dimensional non-Hermitian states with unique bulk, surface, hinge or corner dynamics. Through systematically breaking or restoring $\mathcal{PT}$ symmetry in different sectors of a system, we can selectively activate or manipulate the non-Hermitian skin effect (NHSE) in both the bulk and topological boundary states. Some fascinating phenomena include the directional toggling of the NHSE, and the flow of boundary states without chiral or dynamical pumping, developed from selective boundary NHSE. Our results extend richly into 3D or higher, with more sophisticated interplay with selective bulk and boundary NHSE and charge-parity ($\mathcal{CP}$) symmetry. Based on non-interacting lattices, $\mathcal{PT}$-activated NHSEs can be observed in various optical, photonic, electric and quantum platforms that admit gain/loss and non-reciprocity. 
}
\end{abstract}

\maketitle
\clearpage
\
\newpage
\
\newpage

\maketitle

\noindent
\section*{Introduction}
Parity-time ($\mathcal{PT}$) symmetry is a key ingredient in 
non-Hermitian physics~\cite{PhysRevLett.80.5243,Bender_2007,RevModPhys.93.015005}.
As a special type of pseudo-Hermiticity~\cite{mostafazadeh2002pseudo}, $\mathcal{PT}$ symmetry guarantees real eigenenergies for $\mathcal{PT}$-unbroken eigenstates
, enforcing a balance between gain/loss and thus probability conservation~\cite{PhysRevLett.80.5243,El-Ganainy2018,feng2017non}. As such, it provides an ubiquitous route towards realizing exotic non-Hermitian phenomena in a stable manner, as demonstrated in 
optical~\cite{regensburger2012parity,feng2017non,zhao2018parity,ozdemir2019parity},
acoustic~\cite{zhu2014p,fleury2015invisible,shi2016accessing,shao2020non}, circuit~\cite{choi2018observation,wang2020observation,stegmaier2021topological} and atomic platforms~\cite{hang2013p,zhang2016observation,peng2016anti,jahromi2017statistical,Jiaming2019gainloss,muniz20192d}. 

Recently, it was found that non-Hermitian lattices can possess robust localized modes not just due to topological protection, but also from the non-Hermitian skin effect (NHSE), where all eigenstates accumulate exponentially to the boundary under open boundary conditions (OBCs)~\cite{
lin2023topological,zhang2022review,PhysRevLett.121.086803}. 
In particular, NHSE always requires a complex spectrum under periodic boundary conditions (PBCs) that forms loops in the complex energy plane \cite{PhysRevLett.124.086801,borgnia2020nonH,PhysRevLett.125.126402,PhysRevB.99.201103,li2021quantized}, and hence broken $\mathcal{PT}$ symmetry. 
However, the potential application of $\mathcal{PT}$ symmetry in manipulating NHSE still remains largely unexplored and a comprehensive framework is lacking.

In this work, we found that beyond ensuring stability, $\mathcal{PT}$ symmetry can also serve as a paradigm for designing rich families of higher-dimensional non-Hermitian lattices with unconventional bulk, surface, hinge or corner dynamics.
Note that in our discussion we use ``bulk/surface/edge/hinge'' to describe the co-dimensionality of states in their adiabatically connected Hermitian limit, even if the NHSE has effectively dimensionally-reduced them.
Taking the spectral feature of NHSE as a cornerstone, we can generate various classes of higher-dimensional NHSEs not by breaking $\mathcal{PT}$ symmetry globally, but by selectively activating or breaking it in different sectors e.g. edges and surfaces. 
By systematically switching on/off various types of NHSE in two-dimensional (2D) lattices as examples, we not only obtain rich families of non-Hermitian skin states beyond known NHSEs (e.g. corner NHSE and hybrid skin-topological effect), but also provide a scheme to manipulate them as desired.
In three dimension (3D) or higher,
$\mathcal{PT}$ symmetry can even conspire with charge-parity ($\mathcal{CP}$) symmetry and selective NHSE to activate much richer arrays of non-Hermitian
skin phenomena,
which may prove useful for designing an abundant variety of non-Hermitian optical devices that exploit NHSE~\cite{lasing1,lasing2,lasing3,lasing_review}.
We emphasize that these intriguing models can be realized in classical system, such as electrical circuits~\cite{hofmann2019chiral,li2019emergence,helbig2020generalized,stegmaier2021topological,Zou2021,shang2022experimental,wu2022non,lenggenhager2022simulating,zhu2023higher,PhysRevB.107.085426,2211.09152}.
Moreover, it may also be extended to the quantum systems such as cold atoms and superconducting circuit setups, which are described by Lindblad master equation, where short-time dynamics can be effectively described by non-Hermitian Hamiltonians with the effects of quantum jump ignored~\cite{PhysRevA.100.062131,yi2001effective}.


\section*{Results}
\noindent
\textbf{Corner NHSE from generalized $\mathcal{PT}$ symmetry breaking.}
\begin{figure}
	\includegraphics[width=1\columnwidth]{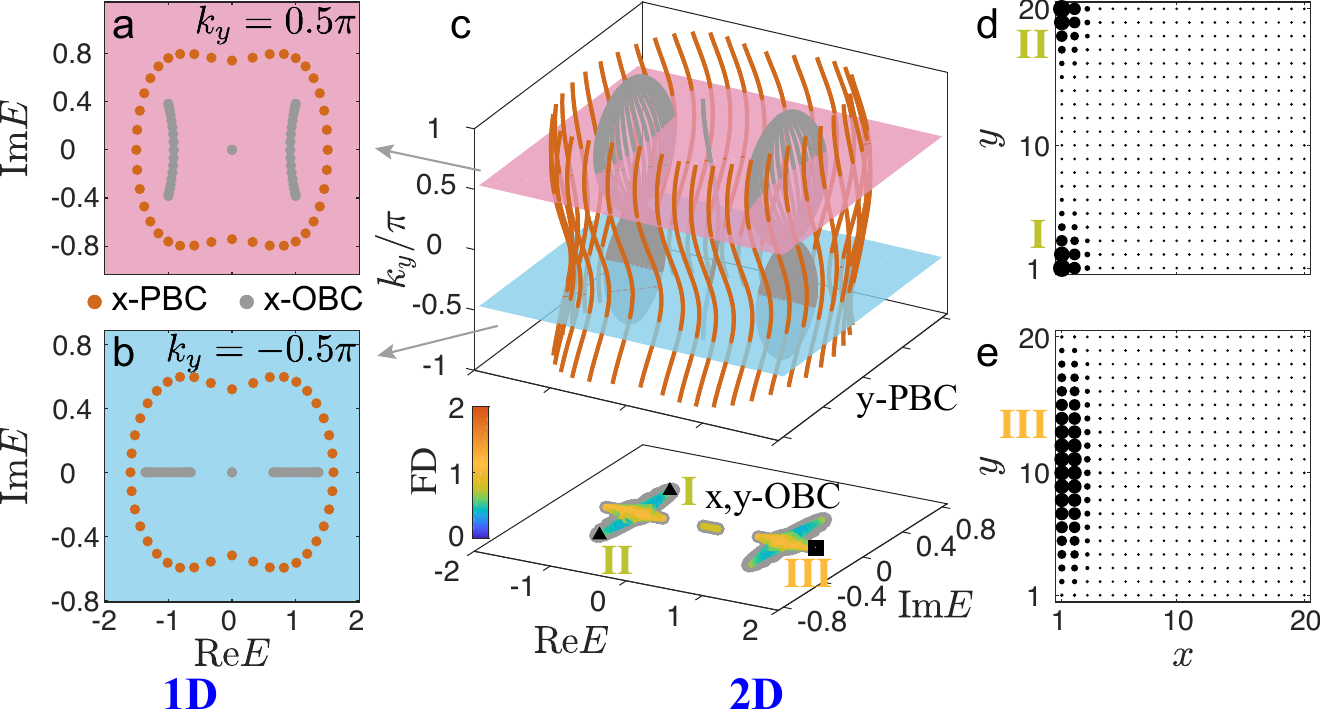}
	\caption{
	\label{fig1_corner} \textbf{Corner and edge modes activated by parity-time ($\mathcal{PT}$) symmetry breaking.} 
	\textbf{a},\textbf{b}: The spectrum under open boundary conditions (OBCs) (gray) of $H_{\gamma,\rm 2D}$ with fixed $k_y$
	can be \textbf{a} complex or \textbf{b} real depending on whether $\mathcal{PT}$ symmetry is restored by the non-Hermitian skin effect. For reference, the spectrum under periodic boundary conditions (PBCs) (brown) is always complex.	
	\textbf{c} Full spectrum of $H_{\gamma,\rm 2D}$ as a 2D model, with PBCs along the $y$ direction. Pink and blue cross sections correspond to the spectra in \textbf{a} and \textbf{b} (also indicated by gray arrows). 
The full $x,y$-OBC spectrum (colored according to the fractal dimension FD) is contained within the set of $y$-PBC eigenenergies, as shown by the lower part of the panel (gray). Corner modes with their fractal dimension FD $\approx 0$ (I,II, triangles) are plotted in \textbf{d}, and exist at $\text{Im}(E)\neq 0$ where $\mathcal{PT}$ symmetry is broken. An edge mode with FD $\approx 1$ (III, square) exists at $\text{Im}(E)= 0$, as plotted in \textbf{e}. Parameters are $\gamma=1.4$, $t_1=0.1$, $t_2=0.1$, $v=1$, and $u=0.7$.
	}
\end{figure}
As a warm-up, we begin with a minimal 2D model $H_{\gamma,\rm 2D}$ whose robust corner/edge modes are localized not from higher-order topology, but from the NHSE activated by $\mathcal{PT}$ symmetry breaking. We construct $H_{\gamma,\rm 2D}$ by extending the 1D non-Hermitian Su-Schrieffer-Heeger (nH-SSH) model~\cite{PhysRevLett.42.1698,PhysRevB.22.2099} to 2D with a $k_y$-dependent coupling (also see Methods section):
\begin{eqnarray}
H_{\gamma,\rm 2D}(k_x,k_y)=h_\gamma^{(0)}\,\sigma_0+h_\gamma^{(x)}\sigma_x+h_\gamma^{(y)}\sigma_y
\label{2D_H_asymmetric}
\end{eqnarray}
where $h_\gamma^{(0)}=t_1\cos k_y$, $h_\gamma^{(x)}=u+v\cos k_x$ and $h_\gamma^{(y)}=v\sin k_x+i\gamma/2+it_2\sin k_y$, and $\sigma_{\alpha=x,y,z}$ are the Pauli matrices acting on a pseudospin-1/2 space (e.g. two sublattices). Here $u,v$ and $t_1,t_2$ are hopping parameters along the $x$ and $y$ directions respectively. 

First consider fixed $k_y$, when this Hamiltonian can be viewed as a 1D nH-SSH model with $x$-NHSE~\cite{PhysRevLett.121.086803} (with an extra energy shift of $t_1\cos k_y$). 
The real-space Hamiltonian of this 1D model satisfies a generalized $\mathcal{PT}$ symmetry $\mathcal{K} H_{\gamma,{\rm 2D}}(k_y)\mathcal{K} =H_{\gamma,{\rm 2D}}(k_y)$, as $x$-OBC Hamiltonian $H_{\gamma,{\rm 2D}}$ contains only real matrix elements~\cite{generalizedPT1,generalizedPT2}.
Under $x$-OBCs, the spectrum remains complex for $k_y\in (0,\pi)$ Fig.~\ref{fig1_corner}a], but becomes real in $k_y\in[ -\pi,0]$ [Fig.~\ref{fig1_corner}b], reflecting the broken and unbroken phases of the generalized $\mathcal{PT}$ symmetry, respectively.
More generally, the real OBC spectrum can be associated to the so-called non-Bloch $\mathcal{PT}$ symmetry ~\cite{xiao2021observation,nBPT1,nBPT2} (also see Methods section),
with non-Bloch exceptional points emerging during transition of these two scenarios at $k=0$ and $\pi$~\cite{xiao2021observation}.

In the 2D context, Figs.~\ref{fig1_corner}a,b correspond to 1D slices [pink and blue planes in Fig.~\ref{fig1_corner}c] of $H_{\gamma,\rm 2D}$ at different quasi-momenta $k_y$, respectively with unbroken/broken generalized $\mathcal{PT}$ symmetry and different types of NHSE. 
When OBCs are also implemented in the $y$-direction, $k_y$ is no longer diagonal and the spectrum with both $x,y$-OBCs [Fig.~\ref{fig1_corner}c Bottom] may differ from the $x$-OBC, $y$-PBC spectrum (Top). In particular, $y$-NHSE can occur only for the states with the generalized $\mathcal{PT}$ symmetry already broken, since the NHSE requires nontrivial spectral winding, which in turn requires complex eigenenergies. 
Since $x$-NHSE acts on all eigenstates of $H_{\gamma,\rm 2D}$, there are no completely extended states, but only edge or corner-localized states under full $x,y$-OBCs, corresponding to single/double direction NHSE from unbroken/broken generalized $\mathcal{PT}$ symmetry, as shown in Fig.~\ref{fig1_corner}d and e respectively. 
To quantify the effective dimensionality of these eigenstates, we introduce the fractal dimension \cite{Theiler:90}
\begin{eqnarray}
{\rm FD}=-\ln[\sum_{\mathbf{r}} |\psi_{n,\mathbf{r}}|^4]/\ln{\sqrt{N}},
\end{eqnarray}
$N$ being the total number of sites and $\psi_{n,\mathbf{r}}$ being the amplitude of $n$-th eigenstates at $\mathbf{r}$-site. We have $\rm FD\approx 0,1$ or $2$ for corner, edge and 2D bulk states. In Fig.~\ref{fig1_corner}c Bottom, ${\rm FD}\approx 1$ for states with real eigenenergies, indicating their 1D edge-localized nature [Fig.~\ref{fig1_corner}e]. In contrast, states with complex eigenenergies have FD closer to $0$, corresponding to the corner localization in Fig.~\ref{fig1_corner}d. 
\\

\noindent\textbf{Directional toggling of NHSE and symmetry-driven corner modes.}
\begin{figure}
	\includegraphics[width=.9\columnwidth]{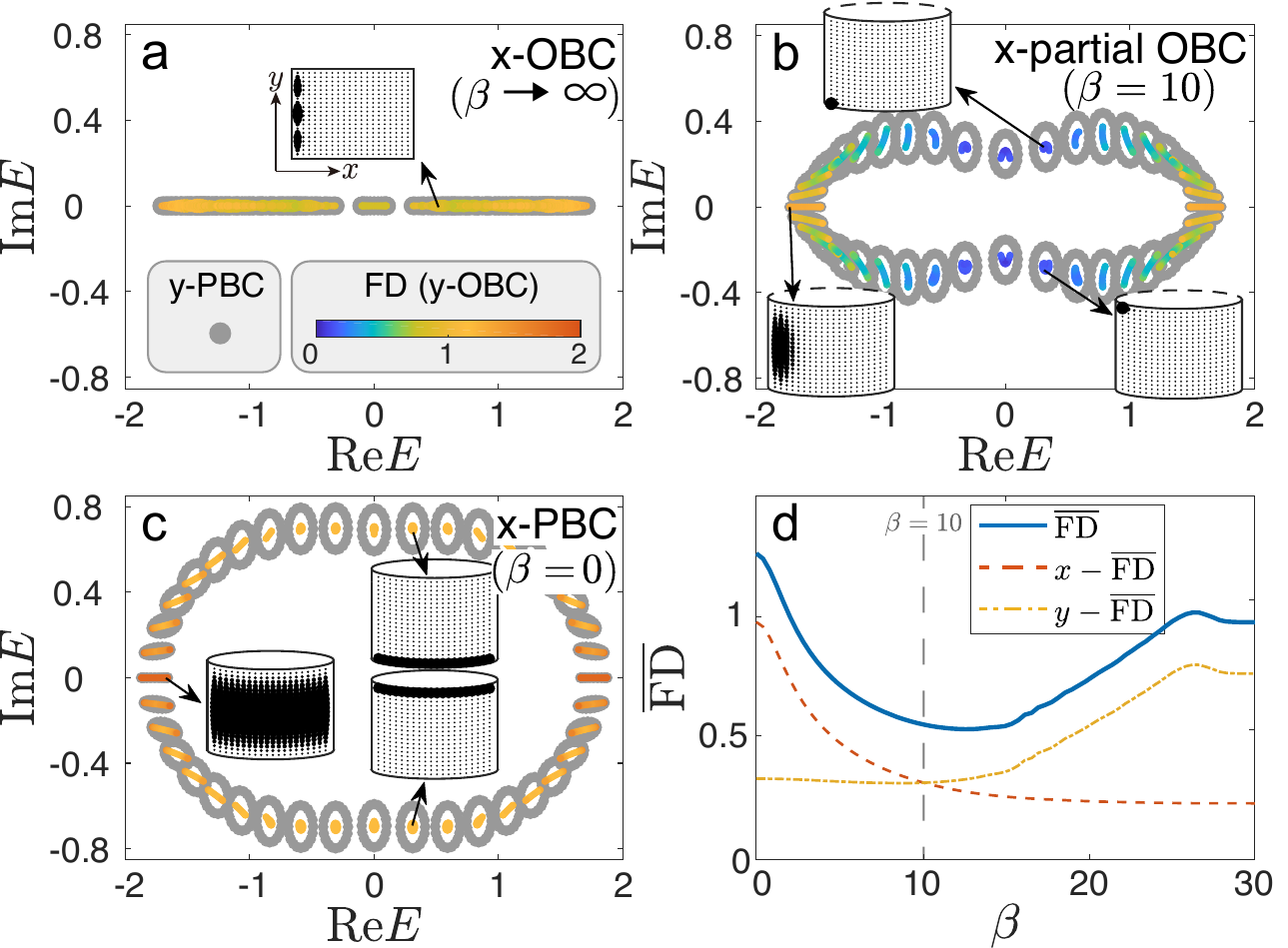}
	\caption{\label{fig2}
	\textbf{Toggling between non-Hermitian skin effect along $x$ and $y$ directions through $x$-boundary tuning.} From \textbf{a} to \textbf{c}: As we morph the boundary conditions along $x$ direction from open boundary conditions (OBCs) \textbf{a} to periodic boundary conditions (PBCs) \textbf{c}, the $y$-PBC (gray) and $y$-OBC (colored by fractal dimension FD) spectra for $H^\beta_{\gamma,\rm 2D}$ transition from purely real to complex. Meanwhile, as activated by $\mathcal{PT}$ symmetry breaking, the skin modes transition from $x$ to $y$-edge localized ($\overline{\rm FD}$ $\approx 1$ with $\overline{\rm FD}$ the average FD over all states), as plotted in the insets. Corner skin modes with $\overline{\rm FD}$ $\approx 0.5$ appear in the intermediate $\beta\approx 10$ regime purely as a by-product of this $\mathcal{PT}$ activation mechanism.	\textbf{d} Average FD and its 1D projections as a function of $\beta$ under the $y$-OBC, with the trough at $\beta\approx 10$ indicative of corner localization.
	Parameters are $\gamma=1.4$, $t_1=0.1$, $t_2=0.1$, $v=1$, and $u=0.9$ in all panels.
	}
\end{figure}
Noting that $\mathcal{PT}$ symmetry plays a crucial role in determining the allowed NHSE directions, we now show that it can be used to activate the NHSE exclusively in the $x$ or $y$ directions. We call this directional NHSE toggling. Furthermore, at an intermediate stage of the toggling, localization occurs briefly in both directions, leading to symmmetry-driven corner modes. 

To demonstrate this intriguing directional toggling of the NHSE, we tune the $\mathcal{PT}$ symmetry by varying the boundary hoppings. We interpolate between $x$-PBCs and $x$-OBCs with a parameter $\beta$, such that $H_{\gamma, \rm 2D}$ becomes
\begin{equation}\label{2D_H_asymmetricIn}
H^\beta_{\gamma,\rm 2D}=H^{\rm OBC}_{\gamma,\rm 2D}+e^{-\beta}H_{1\leftrightarrow N_x}
\end{equation}
where $H^{\rm OBC}_{\gamma,\rm 2D}$ is the Hamiltonian under full $x,y$-OBCs, and $H_{1\leftrightarrow N_x}$ denotes the hoppings between the first and last unit cells along the $x$-direction (see Methods section). 

We have $x$-OBCs and hence $x$-NHSE at $\beta\rightarrow \infty$, and if parameters are chosen such that the generalized $\mathcal{PT}$ symmetry is unbroken for all states, no $y$-NHSE can occur, even though we also have $y$-OBCs. (Corner NHSE shall occur otherwise, see Supplementary Note 1A.)
This gives the $x$-edge-localized modes of Fig.~\ref{fig2}a, with $\rm FD\approx 1$. 
By contrast, the other limit of $\beta=0$ recovers $x$-PBCs 
that eliminate $x$-NHSE, but leads to $y$-NHSE with $y$-localized states [Fig.~\ref{fig2}c]. 
In both scenarios, the $y$-OBCs remain unchanged, yet by adjusting the $x$-boundary hoppings, we can activate the localization in $y$-direction through $\mathcal{PT}$ symmetry breaking.

Interestingly, with partial $x$-OBCs i.e. $\beta=10$, corner modes can appear as an intermediate between $x$ and $y$-boundary localization [Fig.~\ref{fig2}b]. 
These corner modes are evidently distinct from known higher-order skin, topological or hybrid modes, which require proper full OBCs, and may in fact harbor enigmatic scale-free properties inherited from 1D partial boundary states~\cite{li2021impurity,guo2021exact}. We note that some eigenstates always exist at real energies, and are thus always free from the $y$-NHSE.

To quantify the transition between skin localizations along the two directions,
we present in Fig.~\ref{fig2}d
the average fractal dimension $\overline{\rm FD}$ and its 1D projections ($x$-$\overline{\rm FD}$ and $y$-$\overline{\rm FD}$) over all states as functions of $\beta$, where 
\begin{eqnarray}
	&&\overline{\rm FD}=-\sum_{n}\ln[\sum_{\mathbf{r}} |\psi_{n,\mathbf{r}}|^4]/(N\ln{\sqrt{N}}),\nonumber\\
	&&\alpha\text{-}{\rm FD}_n=-\ln[\sum_{\alpha} |\sum_{\alpha'}|\psi_{n,\mathbf{r}}|^2|^2]/\ln{\sqrt{N}},\nonumber\\
	&&\alpha\text{-}\overline{\rm FD}=\sum_{n}\alpha\text{-}{\rm FD}_n/N,
\end{eqnarray}
with $\alpha,\alpha' \in (x,y)$ and $\alpha\neq\alpha'$.
Similar to the  $\overline{\rm FD}$, $\alpha$-$\overline{\rm FD}$ quantifies the locality along a single direction, which is $1$ ($0$) for fully extended (localized) states along the $\alpha$ direction.
While $\beta=0$ and $\beta\gg 1$ gives $\overline{\rm FD}\approx1$, a trough of $\overline{\rm FD}$ exists between them, indicative of corner localization during the transition between single-directional $y$-NHSE under $x$-PBC ($\beta=0$) and $x$-NHSE under $x$-OBC ($\beta\rightarrow \infty$).
Meanwhile, $x$-$\overline{\rm FD}$ decreases and $y$-$\overline{\rm FD}$ increases (almost) monotonously with $\beta$, reflecting the mutual exclusion of NHSE along $x$ and $y$ directions.
\\

\noindent\textbf{Edge-$\mathcal{PT}$-breaking and selective boundary NHSE.}
\begin{figure}
	\includegraphics[width=1\columnwidth]{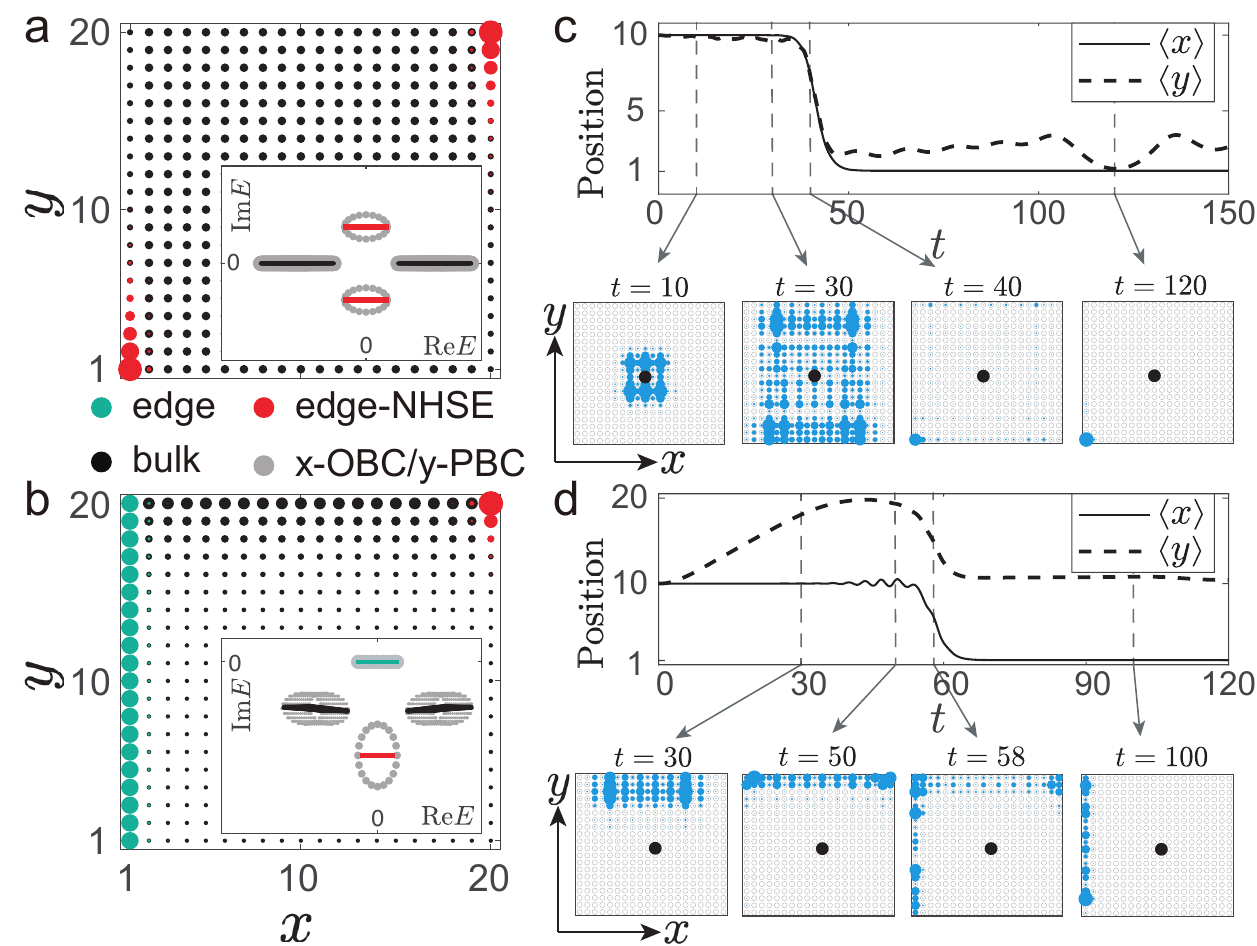}
	\caption{\label{fig3_2D_gainloss}
	\textbf{Selective boundary non-Hermitian skin effect (NHSE) and non-monotonic state dynamics.} 
	\textbf{a} $H_{g,\rm 2D}$ hosts bulk states with unbroken parity-time ($\mathcal{PT}$) symmetry (black) and $\mathcal{PT}$-broken topological edge states under open/periodic boundary conditions (OBCs/PBCs) along $x$/$y$ direction (gray). Under full OBCs, the latter experiences $y$-NHSE and become corner-localized (red). 
	\textbf{b} $H'_{g,\rm 2D}$ is deformed from $H_{g,\rm 2D}$ such that $\mathcal{PT}$ symmetry is restored for the branch of edge states (green) with $\text{Im}(E)>0$, which remains edge-localized; and broken for the bulk states (black), which are thus edge-localized due to $y$-NHSE. 
	\textbf{c},\textbf{d} depict the dynamical evolution of a center-localized initial state (black dot) due to $H_{g,\rm 2D}$ and $H'_{g,\rm 2D}$ respectively, with the size of blue dots indicating evolved state density. Qualitatively distinct stages occur before and after encountering the upper boundary; in \textbf{d}, the upper edge state evolves into the left edge state, leading to non-monotonic $\langle y\rangle$. Parameters are $u=0.2$, $g=0.3$, $t_1=0.3$ and $t'_2=0.1$, with system size $N_x=N_y=20$.
		}
\end{figure}
So far, we have only seen how generalized $\mathcal{PT}$ symmetry breaking can activate various NHSE channels for bulk states. We next discuss how that can interplay with nontrivial topology, which separately gives $\mathcal{PT}$-asymmetric topological edge modes. Specifically, we show that $\mathcal{PT}$ can be selectively broken for edge states only, leading to a type of boundary NHSE distinguished from the usual bulk NHSE. We introduce another 2D model
\begin{eqnarray}
H_{g,\rm 2D}=\sum_{\alpha=0,x,y,z}{h}_g^{(\alpha)}(k_x,k_y)\sigma_\alpha,
\label{eq:H_g}
\end{eqnarray}
where $h_g^{(0)}=t_1\cos k_y$, $h_g^{(x)}=u+v\cos k_x$, $h_g^{(y)}=v\sin k_x$, and imaginary $h_g^{(z)}=i(g+t'_2\sin k_y)$.
This Hamiltonian satisfies the $\mathcal{PT}$ symmetry $[\mathcal{PT},H_{g,\rm 2D}]=0$ with $\mathcal{PT}=\sigma_x\mathcal{K}$, where $\mathcal{K}$ is the complex conjugate operator.
When $u^2+v^2-2uv>g+t_2'$ (all parameters chosen to be positive), all bulk states are $\mathcal{PT}$ unbroken and are free of any type of NHSE.
On the other hand, edge states along $x$-boundaries will appear for all values of $k_y$ when their associating  Wilson loop spectra take nontrivial values~\cite{PhysRevLett.107.036601}, and
the effective edge Hamiltonian read~\cite{PhysRevB.97.205135,PhysRevX.9.011012,PhysRevB.106.245105,PhysRevB.107.115166}:
$$H_{{\rm 2D,edge}}^{\pm}=\hat{P}_{\pm}H_{g,\rm 2D}\hat{P}_{\pm}=t_1\cos k_y\pm i(g+t_2'\sin k_y)$$
with $\hat{P}_{\pm}=(1\pm\sigma_z)/2$ the projectors of edge states.
It is clear that $H_{{\rm 2D,edge}}^{\pm}$ holds no $\mathcal{PT}$ symmetry, and gives a complex edge spectrum [gray dotted loop in the inset of Fig.~\ref{fig3_2D_gainloss}a] with nontrivial spectral winding for $k_y\in (-\pi,\pi]$. 
Consequently, $y$-NHSE occurs for these $x$-boundary states under $y$-OBCs, accumulating on opposite corners (red) depending on the sign of $\text{Im}(E)$ [Fig.~\ref{fig3_2D_gainloss}a].
This mechanism represents a type of hybrid skin-topological effect, namely that topological states are pushed to lower-dimensional boundaries by NHSE, yet bulk states remain extended~\cite{PhysRevLett.123.016805}.
In addition, the $\mathcal{PT}$-protected bulk spectrum remains purely real [Fig.~\ref{fig3_2D_gainloss}a] and impervious to any NHSE throughout.
Indeed, this absence of bulk NHSE is geometry-independent~\cite{Zhang2022,PhysRevLett.131.076401}, representing an intrinsic hybrid skin-topological effect (Supplementary Note 2).

Going further, we may also selectively recover/break $\mathcal{PT}$ symmetry and hence turn off/on the NHSE for different branches of edge states. Explicitly, by introducing an extra anti-Hermitian term $-i (t_2'\sin k_y+g) \sigma_0$:
\begin{eqnarray}
H_{g,\rm 2D}'=H_{g,\rm 2D}-i (t_2'\sin k_y+g) \sigma_0,\label{eq:H_g_2}
\end{eqnarray}
one branch of edge eigenenergies becomes real, $H_{{\rm 2D,edge}}^{+}\rightarrow H_{{\rm 2D,edge}}^{'+}=t_1\cos k_y$.
As seen in Fig.~\ref{fig3_2D_gainloss}b, these edge states remain left edge-localized (green) under full OBCs because they cannot undergo additional NHSE, but those of the other branch are $\mathcal{PT}$-broken and collapse into corner modes (red) due to $y$-NHSE.
\\

\noindent\textbf{Anomalous $\mathcal{PT}$-activated state dynamics.}
The selective activation of these various forms of corner and edge NHSE entails a competition between different NHSE channels, and gives rise to qualitative transitions in different stages of the state dynamics, distinguished to that of different bulk NHSE channels in $H_{\gamma,{\rm 2D}}$ (Supplementary Note 1B). 
Consider an initial state $|\psi(0)\rangle=\left(|\uparrow,N_x/2,N_y/2\rangle+|\downarrow,N_x/2,N_y/2\rangle\right)/\sqrt{2}$ localized at 
the center of an $N_x\times N_y$ lattice. 
We dynamically evolve it via $|\psi(t)\rangle=e^{-iHt}|\psi(0)\rangle$ and investigate the evolution of its center-of-mass $\langle x\rangle$ and $\langle y\rangle$ in Figs.~\ref{fig3_2D_gainloss}c and d, for $H=H_{g,{\rm 2D}}$ and $H'_{g,{\rm 2D}}$ from Eqs. \eqref{eq:H_g} and \eqref{eq:H_g_2} respectively.

For $H=H_{g,{\rm 2D}}$ [Fig.~\ref{fig3_2D_gainloss}c], both $\langle x\rangle$ and $\langle y\rangle$ remain roughly constant for a short time, since $|\psi(t)\rangle$ is governed by NHSE-free bulk dynamics [see Fig.~\ref{fig3_2D_gainloss}a] before it encounters a boundary. After $t\approx 30$, the state arrives at the boundary and the intrinsic hybrid skin-topological effect is activated. The dynamics are then dominated by the corner states with $\text{Im}(E)>0$, where $\langle x\rangle,\langle y\rangle\approx 1$.

A unusual observation is that $\langle x\rangle$, $\langle y\rangle$ can evolve non-monotonically 
in the dynamics governed by $H'_{g,{\rm 2D}}$. 
Due to bulk $y$-NHSE, the state is initially symmetrically pushed to the top boundary, and $\langle y\rangle$ increases while $\langle x\rangle$ remains constant [Fig.~\ref{fig3_2D_gainloss}d]. After 
spreading to the left and right boundaries at $t\approx 50$, the dynamics become dominated by the left edge states with $\text{Im}(E)>0$, and the $y$-NHSE is effectively deactivated. As such, the edge state evolves from the top to the left boundary, and $\langle x\rangle\rightarrow 1$ and $\langle y\rangle\rightarrow N_y/2$. This enigmatic behavior is possible because the left edge states are $\mathcal{PT}$-protected and do not undergo any NHSE to become corner states.
\\

\noindent\textbf{3D generalizations of $\mathcal{PT}$-activated NHSE.}
\begin{figure*}
	\includegraphics[width=\linewidth]{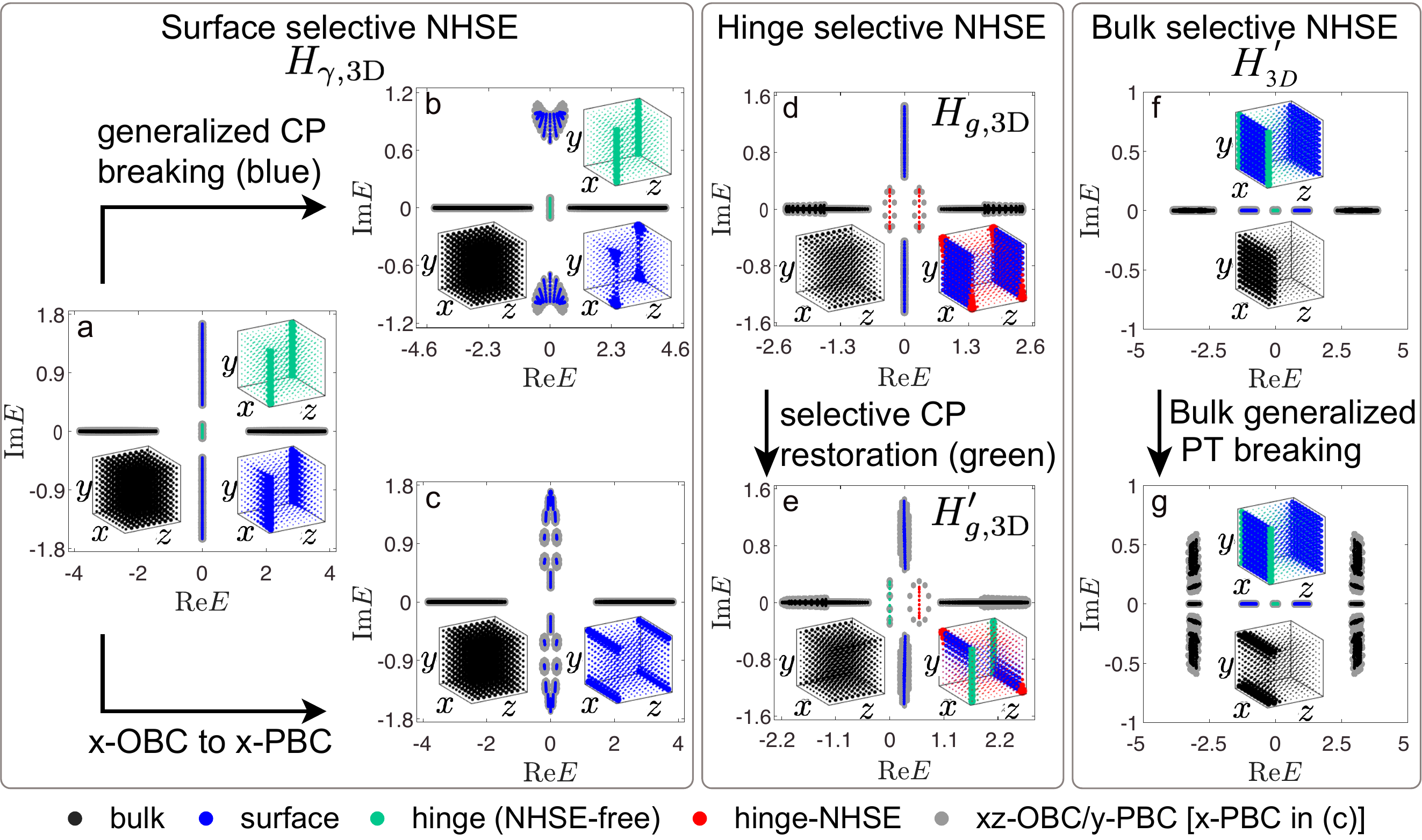}
	\caption{\label{fig3_3Dmain} 
	\textbf{Non-Hermitian skin effect (NHSE) selectively activated by parity-time ($\mathcal{PT}$) or charge-conjugation-parity ($\mathcal{CP}$) symmetry breaking for three-dimensional (3D) models.}
	In most panels [except for \textbf{c} with periodic boundary conditions (PBCs) along $x$], colored dots represent the bulk, surface or hinge spectra under full OBCs, and gray dots represent the spectra with $xz$-OBCs and PBCs along $y$ direction. Insets of 3D cubes display the average spatial distribution of eigenstates of the respective color. \textbf{a}-\textbf{c}, \textbf{d}, \textbf{e}, and \textbf{f}-\textbf{g} pertain to $H_{\gamma,\text{3D}}$, $H_{g,\text{3D}}$, $H'_{g,\text{3D}}$, and $H'_\text{3D}$ respectively. In \textbf{a},\textbf{b},\textbf{c}, $(001)$ and $(00\bar{1})$-surface states (blue) experience corner pumping from $\mathcal{CP}$-breaking \textbf{b} or $x$ to $y$-NHSE directional toggling \textbf{a},\textbf{c}. In \textbf{d},\textbf{e}, hinge states (red) can have $\mathcal{CP}$ symmetry selectively broken (unbroken), respectively resulting in corner (red) or hinge (green) localization. In \textbf{f},\textbf{g}, on-demand switching of 1st and 2nd-order 3D bulk NHSE (black surface and edges) through $\mathcal{PT}$ breaking leaves other states (green,blue) invariant.
		In \textbf{a} to \textbf{c}, parameters are $v=1$, $u'=0.6$, $v'=3$ and $t_1=t_2=0.1$, with $u=0.6$ and $\gamma=0.4$ for \textbf{a} and \textbf{c}; $u=0.5$ and $\gamma=1.4$ for \textbf{b}. 
	In \textbf{d} and \textbf{e}, parameters are $u=0.2$, $v=1$, $u'=0.4$, $v'=2$, $g=0.3$, $t_1=0.3$ and $t_2=0.1$.
	In \textbf{f} and \textbf{g}, parameters are $u'=0.2$, $v'=1$, $v=3$, $\gamma=1.4$ and $t_1=t_2=0.1$, with $u=0.9$ for \textbf{f} and $u=0.5$ for \textbf{g}. 
	}
\end{figure*}
The various avenues for $\mathcal{PT}$-activated NHSE effects discussed above takes on even richer possibilities in higher dimensions. Below, we briefly demonstrate how they acquire extra interpretations as surface and hinge-selective NHSE in 3D. Specifically, we consider a 3D Hamiltonian of the form
\begin{eqnarray}
	&&H_{\gamma/g,\rm 3D}(\mathbf{k})=H_{\gamma/g,0}(k_x,k_y)\nonumber\\
	&&~~~~~~~~~~+(u'+v'\cos k_z)\sigma_0\tau_x+v'\sin k_z\sigma_0\tau_y	
	\label{eq:3D}
\end{eqnarray}
with $\tau_{x,y,z}$ another set of Pauli matrices. $H_{\gamma/g,0}(k_x,k_y)$ is inherited from the previous $\mathcal{PT}$-breaking and directional toggling NHSE models $H_{\gamma/g,\rm{2D}}$:
\begin{eqnarray}
	H_{\gamma,0}=iH_{\gamma,\rm 2D}\oplus \left(-iH^\dagger_{\gamma,\rm 2D}\right),~
	H_{g,0}=iH_{g,\rm 2D}\tau_z,\label{eq:3D_2D}
\end{eqnarray}
which contain only diagonal terms of $\tau_{0,z}$.
Since $k_z$ enters only off-diagonal terms of $\tau_{x,y}$,
the surfaces states under $z$-OBCs obey the effective Hamiltonian
$H_{\pm,{\rm surface}}=\hat{P}_{\pm,{\rm 3D}}H_{\gamma/g,\rm 3D}\hat{P}_{\pm,{\rm 3D}}$ with $\hat{P}_{\pm,{\rm 3D}}=(1\pm \sigma_0\tau_z)/2$ $z$-surface state projectors~\cite{PhysRevB.97.205135,PhysRevX.9.011012,PhysRevB.106.245105,PhysRevB.107.115166}.  
Note that the choices $H_{\gamma/g,0}$ are not unique, and the explicit forms in Eq. \eqref{eq:3D_2D} are chosen to
restore $\mathcal{PT}$ symmetry 
\begin{eqnarray}
	[\sigma_x\tau_x\mathcal{K},H_{\gamma/g,\rm 3D}(\mathbf{k})]=0.\label{eq:PT_3D} 
\end{eqnarray}
Also, an additional $\mathcal{CP}$ symmetry emerges for the surface states~\cite{PhysRevX.8.031079} (also see Methods section), which offers another avenue for eliminating the NHSE by ensuring purely imaginary eigenenergies.

In Fig.~\ref{fig3_3Dmain}, eigenmodes 
are marked with different colors according to their corresponding sectors and boundary conditions.
In Fig.~\ref{fig3_3Dmain}a,b,c for $H_{\gamma,\rm 3D}$, the $(001)$ and ($00\bar{1}$) surface states essentially recapitulate the symmetry-selected and directional toggling NHSE effects of Figs.~\ref{fig1_corner} and \ref{fig2}. Going from Fig.~\ref{fig3_3Dmain}a to b, these surface states (blue) become corner-localized due to generalized $\mathcal{CP}$-breaking, analogous to the corner localization Fig.~\ref{fig1_corner}d due to $\mathcal{PT}$-breaking, while bulk (black) and hinge (green) states remain unchanged. Likewise, directional toggling from $x$-NHSE to $y$-NHSE localization occurs when the $x$-OBC of Fig.~\ref{fig3_3Dmain}a is replaced by the $x$-PBC of Fig.~\ref{fig3_3Dmain}c.

Similarly, the surface (blue) and hinge (red) states on $(001)$ and ($00\bar{1}$) surfaces of $H_{g,\rm 3D}$ in Fig.~\ref{fig3_3Dmain}d are in direct correspondence with the bulk (black) and boundary (red) states of $H_{g,{\rm 2D}}$ in Fig.~\ref{fig3_2D_gainloss}a.
Meanwhile, bulk eigenenergies remain real as our choice of parameters maintains unbroken $\mathcal{PT}$ symmetry in Eq.~\eqref{eq:3D}.  
In analogy with $H'_{g,{\rm 2D}}$ in Fig.~\ref{fig3_2D_gainloss}b, 
a $\mathcal{CP}$ symmetry can be recovered in
one branch of hinge states on each of $(001)$ and ($00\bar{1}$) surfaces by introducing an extra Hermitian term to the Hamiltonian, i.e. $H'_{g,\rm 3D}(\mathbf{k})=H_{g,\rm 3D}(\mathbf{k})+(t_2'\sin k_y+g)\sigma_0\tau_0$, which removes the NHSE on these hinge states [Fig.~\ref{fig3_3Dmain}e].


Alternatively, a 3D Hamiltonian with only bulk-NHSE 
can be designed such that its bulk (PBC) states are not $\mathcal{PT}$-symmetric, but that generalized $\mathcal{PT}$-symmetry can be recovered in certain parameter regimes, 
for instance $H'_{\rm 3D}(\mathbf{k})=H_{\gamma,\rm{2D}}(k_z,k_y)\otimes\tau_0+\sigma_z\otimes H_{\rm SSH}(k_x)$,
with
\begin{eqnarray}
	H_{{\rm SSH}}(k_x)&=&(u'+v'\cos k_x)\tau_x+v'\sin k_x\tau_y	
	\label{sub_3D}
\end{eqnarray} 
the Hermitian SSH model and $H_{\gamma,\rm{2D}}(k_z,k_y)$ from Eq.~\eqref{2D_H_asymmetric}.
Consequently, bulk states accumulate to a 2D surface (or 1D hinges) when the generalized $\mathcal{PT}$ symmetry is unbroken (or broken), as shown in Fig.~\ref{fig3_3Dmain}f,g. In contrast to $H_{\gamma/g,\text{3D}}$, conventional surface (blue) or hinge (green) states are immune from the NHSE and remain extended. 
Thereby, our work provides a systemic framework for constructing and exploring various types of $\mathcal{PT}$ activated bulk and boundary skin effect, which also can be applicable in the study of non-Hermitian gapless phases\cite{PhysRevB.104.L161116,PhysRevB.104.L161117}.

\section*{Discussions}
\noindent
$\mathcal{PT}$ symmetry ensures real spectra, thus eliminating the prospect for spectral winding. This fundamental observation leads to the  paradigm of $\mathcal{PT}$-activated NHSE, where rich families of NHSE-related phenomena can be designed by selectively symmetry breaking in bulk or boundary subspaces. In 2D, we discussed the directional NHSE toggling and $\mathcal{PT}$-mediated corner modes, as well as selective edge-$\mathcal{PT}$ breaking and restoring that causes non-monotonic transfer of edge states. In 3D or higher, $\mathcal{PT}$ activation leads to far more varied phenomenology by interplaying with the already rich classes of boundary and bulk NHSE. This paradigm of symmetry-controlled NHSE can be extended to include other symmetries which forbid spectral winding, such as non-Bloch and generalized $\mathcal{PT}$ symmetries, $\mathcal{CP}$ symmetry, or pseudo-Hermiticity. 
It also provides a versatile scheme for constructing models with different types of NHSE on different sectors of the systems.
We also note that the conditions for yielding a real OBC spectrum with complex PBC spectrum can be geometrically understood with an electrostatics analogy of the NHSE problem~\cite{electrostatics}, which may provide further possibilities for activating/deactivating NHSE based on our method.

Since $\mathcal{PT}$-activated NHSE is fundamentally a phenomenon on a linear lattice, it can be experimentally demonstrated on any metamaterial or artificial lattice, whether classical or quantum, as long as $\mathcal{PT}$ symmetry can be broken through a combination of gain/loss and non-reciprocity. Possible platforms include non-reciprocal lossy acoustic or photonic crystals or waveguides~\cite{xiao2020non,lin2022observation,xiao2023observation,wang2023experimental}, ultracold atomic lattices~\cite{PhysRevLett.129.070401}, superconducting quantum circuit lattices~\cite{kollar2019hyperbolic,smith2019simulating,bassman2021simulating,ippoliti2021many,koh2022stabilizing,koh2022simulation,zhang2022digital,koh2023observation,fleckenstein2022non,frey2022realization,chen2022high}, and in particular op-amp controlled electrical circuits which can be constructed with great versatility~\cite{hofmann2019chiral,li2019emergence,helbig2020generalized,stegmaier2021topological,Zou2021,shang2022experimental,wu2022non,lenggenhager2022simulating,zhu2023higher,PhysRevB.107.085426,2211.09152}. 

\section*{Methods}
\begin{figure}
	\includegraphics[width=1\columnwidth]{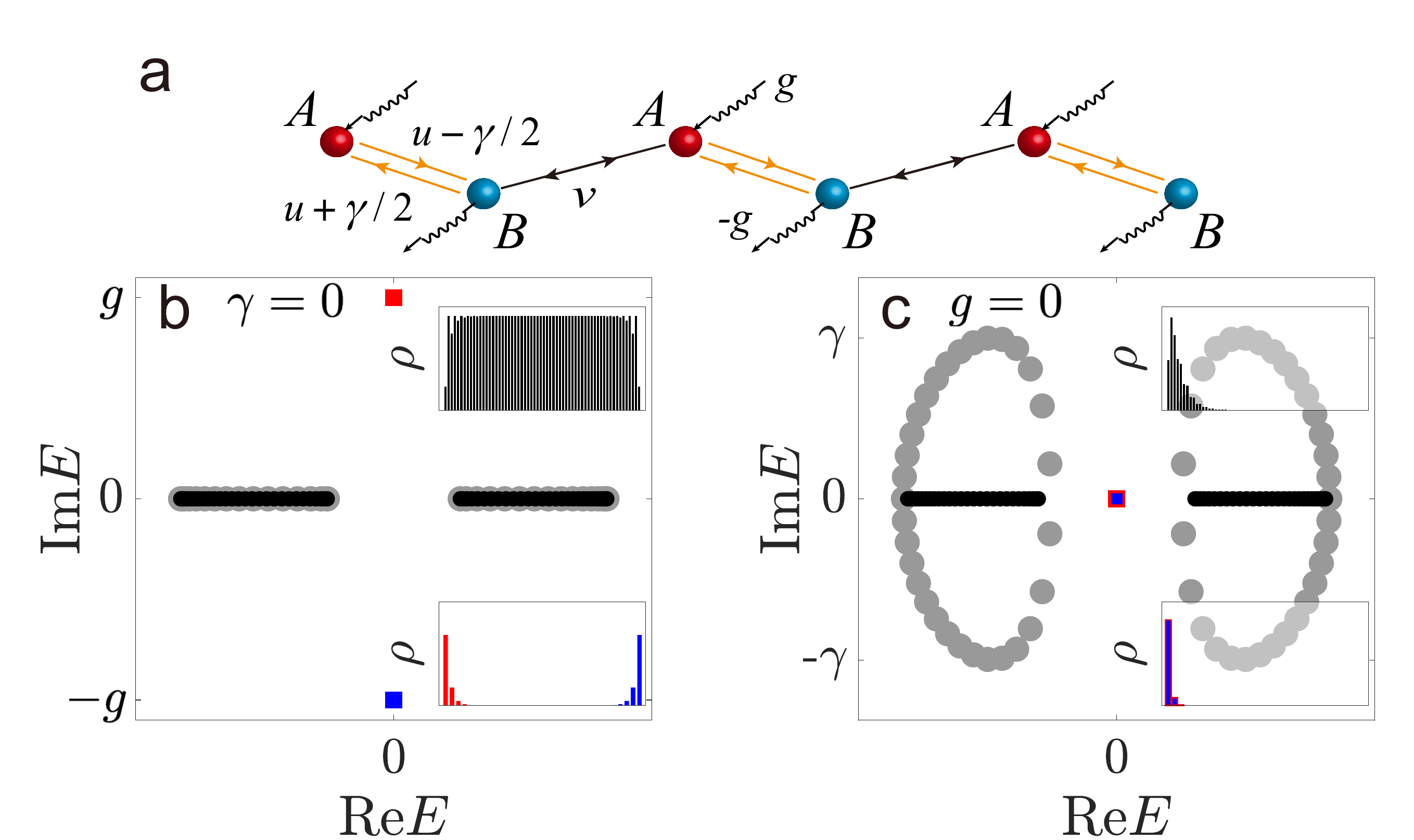}
	\caption{\label{supfig1_SSH}
		\textbf{The non-Hermitian Su-Schrieffer-Heeger  model.} \textbf{a} A sketch of the model of Eq. \eqref{1D_SSH}, with $\gamma$ ($g$) the asymmetric hopping (gain/loss) parameter.
		\textbf{b} Typical spectra  of $H_g$ under periodic (gray) and open (other colors) boundary conditions (PBCs and OBCs).
		Under the OBCs, bulk states can be parity-time-symmetric ($\mathcal{PT}$ symmetric) and have real eigenenergies (black), but edge states are $\mathcal{PT}$-broken and have imaginary eigenenergies $\pm i g$ (red and blue).
		Insets demonstrate the spatial distribution of corresponding eigenstates.
		\textbf{c} Typical PBC and OBC spectra of $H_\gamma$, with the same color marks as in \textbf{b}. The PBC spectrum in \textbf{c} is $\mathcal{PT}$-broken and possesses a nontrivial spectral winding, leading to a boundary accumulation of all eigenstates (see insets). The OBC spectrum restores a non Bloch (or a generalized) $\mathcal{PT}$ symmetry and becomes purely real.
		Parameters are chosen to be $u=0.5$, $v=1$, and \textbf{b} $g=0.2$, \textbf{c} $\gamma=0.2$ for demonstration.
	}
\end{figure}

\noindent
\textbf{Non-Hermitian SSH model}\label{appA}
The models supporting various types of NHSE discussed in the main text are mostly designed based on a non-Hermitian Su-Schrieffer-Heeger (SSH) model~\cite{PhysRevLett.42.1698,PhysRevB.22.2099}, described by the Hamiltonian:
\begin{eqnarray}\label{1D_SSH}
	H(k_x)=\sum_{\alpha=x,y,z}{h}_\alpha(k_x)\sigma_\alpha,
\end{eqnarray}
where $h_x(k_x)=u+v\cos k_x$, $h_y(k_x)=v\sin k_x+i\gamma/2$, $h_z=ig$,
and $\sigma_{\alpha=x,y,z}$ is the Pauli matrix acting on a pseudospin-1/2 space [e.g. two sublattices $|A\rangle$ and $|B\rangle$ as shown in Fig. \ref{supfig1_SSH}a].
$u$ and $v$ denote the Hermitian staggered hopping amplitudes.
The two non-Hermitian parameters $\gamma$ and $g$ describes asymmetric hoppings and imaginary on-site potential, which correspond to to non-local and local dissipation~\cite{PhysRevLett.123.170401}, respectively, as illustrated in Fig. \ref{supfig1_SSH}a.
By separately tuning the two non-Hermitian parameters $g$ and $\gamma$, we can break or restore $\mathcal{PT}$-symmetry selectively for bulk or topological edge states in this model, as discussed below.

When $\gamma=0$,
non-Hermiticity enters the Hamiltonian only through the on-site gain and loss of $g$, and the resultant Hamiltonian
$$H_{g}\equiv H(\gamma=0)$$
holds a $\mathcal{PT}$ symmetry: $[\mathcal{PT},H_{g}]=0$ with $\mathcal{PT}=\sigma_x\mathcal{K}$ and $\mathcal{PT}^2=+1$.
A nonzero $g$ does not break the $\mathcal{PT}$ symmetry as the gain and loss are balanced between sublattices. As shown in
Fig. \ref{supfig1_SSH}b, in the $\mathcal{PT}$-unbroken phase, all bulk states have real eigenenergies since they are
also eigenstates of $\mathcal{PT}$ symmetry operator~\cite{PhysRevLett.80.5243,Bender_2007}.
On the other hand, topological edge states for the SSH model are sublattice polarized and hence $\mathcal{PT}$-broken.
Instead of being eigenstates of $\mathcal{PT}$ symmetry operator, these states are related to each other through the symmetry, i.e. $\mathcal{PT}|\psi_+\rangle \propto |\psi_-\rangle$, with $|\psi_{+,-}\rangle$ the topological edge states localized at left and right ends respectively.
The sublattice-polarization of these edge states allow us to write down their effective Hamiltonians through projectors $\hat{P}_{\pm}=(1\pm\sigma_z)/2$,
i.e. $$H_{{\rm edge}}^{\pm}=\hat{P}_{\pm}H_{g}\hat{P}_{\pm}=\pm ig,$$ which directly give their eigenenergies~\cite{PhysRevB.97.205135,PhysRevX.9.011012,PhysRevB.106.245105,PhysRevB.107.115166}.

For the other non-Hermitian scenario of Eq. \eqref{1D_SSH} with $g=0$ but $\gamma\neq 0$,
\begin{eqnarray}\label{1D_SSHgamma}
	H_{\gamma}\equiv H(g=0)
\end{eqnarray}
the $\mathcal{PT}$ symmetry is broken, and the resultant complex bulk spectrum is known to have nontrivial spectral winding topology under the PBC,
a signature of the emergence of NHSE under the OBC \cite{borgnia2020nonH,PhysRevLett.125.126402,PhysRevLett.124.086801}.
Yet its OBC bulk spectrum become real in Fig. \ref{supfig1_SSH}c due to the recovery of a generalized symmetry $\mathcal{PT}$ symmetry $\mathcal{K}H_{\gamma}\mathcal{K}=H_{\gamma}$~\cite{generalizedPT1,generalizedPT2} or a non-Bloch $\mathcal{PT}$ symmetry~\cite{xiao2021observation,nBPT1,nBPT2},
 or more generally, a pseudo-Hermiticity~\cite{mostafazadeh2002pseudo}. Explicitly, the OBC bulk spectrum can be described by a non-Bloch Hamiltonian $\bar{H}_{\gamma}(k_x)=H_{\gamma}(k_x+i\kappa)$ with $\kappa=\ln\sqrt{|(u+\gamma/2)/(u-\gamma/2)|}$ \cite{PhysRevLett.121.086803,yokomizo2019non,PhysRevB.99.201103}.
That is, this non-Hermitian Hamiltonian can be transformed into a Hamiltonian without NHSE through a similarity transformation\cite{PhysRevLett.121.086803}: $\hat{\psi}^{\dag}_{j,A/B}\rightarrow e^{-j\kappa}\hat{\psi}^{\dag}_{j,A/B}$, where $\hat{\psi}^{\dag}_{j,A/B}$ is the annihilate operator for $j$-th cell with sublattices $A/B$. In particular, a Hermitian SSH model will be obtained directly through this transformation in the non-Bloch $\mathcal{PT}$ unbroken phase.
A different route of transforming non-Hermitian Hamiltonians into Hermitian ones can be found in Ref. \cite{PhysRevResearch.4.023070}.
This Hamiltonian $\bar{H}_{\gamma}(k_x)$ satisfies a non-Bloch $\mathcal{PT}$ symmetry,
$$[\eta\mathcal{K},\bar{H}_{\gamma}]=0$$
with $\eta$ a unitary operator, which is unbroken for all non-Bloch eigenstates when $|u|>|\gamma/2|$ (as discussed later).
In contrast, here the topological edge states always remain $\mathcal{PT}$-symmetric and possess real (zero) eigenenergies, as a nonzero $\gamma$ does not enter their effective Hamiltonian: $$H_{{\rm edge}}^{\pm}=\hat{P}_{\pm}H_{\gamma}\hat{P}_{\pm}=0.$$
Note that the the localizing direction of these edge states may be altered by the NHSE, yet their distributing dimensionality remains unchanged, namely they are always 0D edge states except for some critical points in the parameter space \cite{zhu2021delocalization}.
Similarly, edge states of $2$D generalization Hamiltonian $H_{\gamma,\rm 2D}$ with selective bulk NHSE, namely those near zero energy in Fig.1c and Fig.2a of the main text, are seen to also possess real eigenergies with their distributing dimensionality unchanged by NHSE.

\textbf{Non-Bloch $\mathcal{PT}$ symmetry.}
While the models of $H_\gamma$ in the last section and $H_{\gamma, {\rm 2D}}$ satisfy a generalized $\mathcal{PT}$ symmetry,
it is the non-Bloch $\mathcal{PT}$ symmetry that guarantees real OBC spectrum for more generic non-Hermitian Hamiltonians.
Here, we discuss the non-Bloch $\mathcal{PT}$ symmetry of the non-Hermitian SSH model with asymmetric hoppings $$H_{\gamma}(k_x)=(u+v\cos k_x)\sigma_x+(\sin k_x+i\gamma/2)\sigma_y.$$
When $\gamma=0$, this Hamiltonian satisfies the conventional $\mathcal{PT}$ symmetry $[\sigma_x\mathcal{K},H_{\gamma}(k_x)]=0$ with  $\mathcal{K}$ the complex conjugate operator.
Its corresponding non-Bloch Hamiltonian reads~\cite{PhysRevLett.121.086803,yokomizo2019non,PhysRevB.99.201103}
\begin{eqnarray}
	\bar{H}_{\gamma}(k_x)=H_\gamma(k_x+i\kappa)=(A+iB)\sigma_x+(C+iD)\sigma_y,\nonumber\\
\end{eqnarray}
with
\begin{eqnarray}\label{fac}
	A&=&u+v\frac{\cos k_x}{2}(W_{+}+W_{-}),\nonumber\\
	B&=&-v\frac{\sin k_x}{2}(W_{+}-W_{-}),\nonumber\\
	C&=&v\frac{\sin k_x}{2}(W_{+}+W_{-}),\nonumber\\
	D&=&v\frac{\cos k_x}{2}(W_{+}-W_{-})+\frac{\gamma}{2},
\end{eqnarray}
with $W_{\pm}=\sqrt{(u\pm\gamma/2)/(u\mp\gamma/2)}$. 
The quantity $\kappa$ describes the inverse localization length of skin modes under OBCs, which can be obtained by requiring non-Bloch eigenenergies to form degenerate pairs for different quasi-momenta  (so that their linear combination can satisfy OBCs).
$W_{\pm}$ as well as all factors defined in Eq.~\eqref{fac} are real when $|u|>|\gamma/2|$. In this regime, we can define non-Bloch $\mathcal{PT}$ symmetry. To identify the non-Bloch $\mathcal{PT}$ symmetry, we rewrite the non-Bloch Hamiltonian as
\begin{eqnarray}
	H_\gamma(k_x+i\kappa)=h_x'\sigma_x'+h_y'\sigma_y'
\end{eqnarray}
with
\begin{eqnarray}
 h_x'&=&\frac{{(A+iB)(W_++W_-)+i(C+iD)(W_+-W_-)}}{2},\nonumber\\
 h_y'&=&\frac{{(C+iD)}(W_++W_-)-i(A+iB)(W_+-W_-)}{2},\nonumber\\
 \sigma_x'&=&\frac{(W_++W_-)\sigma_x+i(W_+-W_-)\sigma_y}{2},\nonumber\\ \sigma_y'&=&\frac{(W_++W_-)\sigma_y-i(W_+-W_-)\sigma_x}{2}.\nonumber\\
\end{eqnarray}
It is straightforward to check that $h_x',h_y'\in \mathbb{R}$, $(\sigma_x')^2=(\sigma_y')^2=\sigma_0$, and $\sigma_x' \sigma_y'\sigma_x'=-\sigma_y'$.
Therefore we obtain
\begin{eqnarray}\label{comm}
	[\eta \mathcal{K},H_\gamma(k_x+i\kappa)]=0\label{eq:nonB_PT}
\end{eqnarray}
with $\eta=\sigma_x'$ a parameter-dependent unitary operator and $(\eta \mathcal{K})^2=1$, indicating the non-Bloch $\mathcal{PT}$ symmetry analogous to the conventional one.

We emphasis here that the system is non-Bloch $\mathcal{PT}$ symmetric only when $|u|>|\gamma/2|$,
as Eq.~\eqref{comm} is not satisfied when $|u|<|\gamma/2|$. 
In the non-Bloch $\mathcal{PT}$ symmetric regime with $|u|>|\gamma/2|$,
eigenenergies
of $H_\gamma(k_x+i\kappa)$, i.e.
\begin{eqnarray}
	E_\gamma=\pm\sqrt{G+2i(AB+CD)}
\end{eqnarray}
with $G=A^2+C^2-B^2-D^2$,
always take real values, since we have
\begin{eqnarray}
	AB+CD&=&v\frac{\sin k_x}{2}\Delta\sqrt{|u^2-\gamma^2/4|}=0
\end{eqnarray}
with $\Delta={\rm sgn}[u+\gamma/2]-{\rm sgn}[u-\gamma/2]$,
and 
\begin{eqnarray}
G&=&u^2+v^2-\gamma^2/4\nonumber+v\cos k_x
	\Delta\sqrt{|u^2-\gamma^2/4|}\nonumber\\
	&\geq&(\sqrt{u^2-\gamma^2/4}-v)^2\geq0.
\end{eqnarray}
In other words, the system always falls in the non-Bloch $\mathcal{PT}$ unbroken phase in the symmetric regime of $|u|>|\gamma/2|$.
On the other hand, when $|u|<|\gamma/2|$, the system no longer holds the symmetry of Eq.~\eqref{comm}, and the eigenenergies $E_\gamma$ generally become complex. 
An exception is when $k_x\in\{0,\pi\}$, 
where the eigenenergies reduce to
\begin{eqnarray}
	E_\gamma(k=0,\pi)=\pm\sqrt{u^2+v^2-\gamma^2/4}
\end{eqnarray}
when $|u|<|\gamma/2|$,
which are also real provided $u^2+v^2>\gamma^2/4$.

\noindent\textbf{The explicit form of Hamiltonian $H^\beta_{\gamma,\rm 2D}$.}
For analytic tractability, we give the explicit form of $H^\beta_{\gamma,\rm 2D}$ in Eq.~(\ref{2D_H_asymmetricIn}).
According to this definition, the Hamiltonian $H^\beta_{\gamma,\rm 2D}$ contains two parts: the Hamiltonian $H_{\gamma,\rm 2D}$ in Eq.~(\ref{2D_H_asymmetricIn}) under full OBCs, and its hopping between boundaries along $x$-direction.
The first part reads
\bigskip
\begin{widetext}
\begin{eqnarray}\label{OBCHgamma}
	H^{\rm OBC}_{\gamma,\rm 2D}=&&\frac{v}{2}\sum_{j}^{N_x-1}\sum_{k}^{N_y}[\hat{\psi}^{\dag}_{j,k}(\sigma_x-i\sigma_y)\hat{\psi}_{j+1,k}+ {\rm H.c.}]\nonumber \\
	&&+\frac{1}{2}\sum_{j}^{N_x}\sum_{k}^{N_y-1}[t_1(\hat{\psi}^{\dag}_{j,k}\hat{\psi}_{j,k+1}+{\rm H.c.})-it_2(i\hat{\psi}^{\dag}_{j,k}\sigma_y\hat{\psi}_{j,k+1}+{\rm H.c.})]\nonumber \\
	&&+\sum_{j}^{N_x}\sum_{k}^{N_y}\hat{\psi}^{\dag}_{j,k}(u\sigma_x+\frac{i\gamma}{2}\sigma_y)\hat{\psi}_{j,k},
\end{eqnarray}
\end{widetext}
where $\hat{\psi}_{j,k}=[\hat{a}_{j,k},\hat{b}_{j,k}]^T$, with $\hat{a}_{j,k}$ ($\hat{b}_{j,k}$) being the annihilate operator for the pseudospin-$\uparrow$ ($\downarrow$) [sublattices $A$ ($B$)] on the $(j,k)$-th unit cell.

Similarly, the hoppings between boundaries along $x$-direction $H_{1\leftrightarrow N_x}$ and $y$-direction $H_{1\leftrightarrow N_y}$ are
\begin{eqnarray}\label{boundHgamma}
	H_{1\leftrightarrow N_x}=&&\frac{v}{2}\sum_{k}^{N_y}[\hat{\psi}^{\dag}_{N_x,k}(\sigma_x-i\sigma_y)\hat{\psi}_{1,k}+ {\rm H.c.}],\nonumber\\
	H_{1\leftrightarrow N_y}=&&\frac{t_1}{2}\sum_{j}^{N_x}(\hat{\psi}^{\dag}_{j,N_y}\hat{\psi}_{j,1}+{\rm H.c.})\nonumber\\
	&&-\frac{it_2}{2}\sum_{j}^{N_x}(\hat{\psi}^{\dag}_{j,N_y}\sigma_y\hat{\psi}_{j,1}+{\rm H.c.}).
\end{eqnarray}
Thereby, the Hamiltonian under full PBCs can be  expressed by
\begin{eqnarray}\label{PBCHgamma}
	H^{\rm PBC}_{\gamma,\rm 2D}=H^{\rm OBC}_{\gamma,\rm 2D}+H_{1\leftrightarrow N_x}+H_{1\leftrightarrow N_y}.
\end{eqnarray}
Through the Fourier transformation, the momentum space Hamiltonian $H_{\gamma,\rm 2D}(k_x,k_y)$ in Eq.~(\ref{2D_H_asymmetricIn}) can be obtained.
In Eq.~(\ref{2D_H_asymmetricIn}), an extra factor $e^{-\beta}$ is introduced to describe the partial OBCs aong $x$ direction.
\\

\noindent\textbf{Effective surface Hamiltonian for $3$D generalizations.}
The Hamiltonians $H_{\gamma/g,\rm 3D}$ and $H'_{\gamma,\rm 3D}$ are chosen to support surface states on $(001)$ and $(00\bar{1})$ surfaces.
Corresponding projectors of these surface states are $$\hat{P}_{\pm,{\rm 3D}}=(1\pm \sigma_0\tau_z)/2.$$
Thereby, the effective surface Hamiltonian of $H_{\gamma,\rm 3D}$ can be given by:
\begin{eqnarray}
	H^{\pm}_{\gamma,\rm surf}&=&\hat{P}_{\pm,{\rm 3D}}H_{\gamma,\rm 3D}\hat{P}_{\pm,{\rm 3D}}\nonumber\\
	&=&\pm i\{t_1\cos k_y\sigma_0+(u+v\cos k_x)\sigma_x\nonumber\\
	&&+[v\sin k_x\pm i(\gamma/2+t_2\sin k_y)]\sigma_y\}.\label{eq:progmma}
\end{eqnarray}
Comparing to Eq.~(\ref{2D_H_asymmetricIn}), we find the effective Hamiltonian $H^{\pm}_{\gamma,\rm surf}$ can be considered as $\pm i H_{\gamma,\rm 2D}$ (the signs of $\gamma$ and $t_2$ are flipped for $H^{-}_{\gamma,\rm surf}$).
As discussed in before, $H_{\gamma,\rm 2D}$ holds a non-Bloch $\mathcal{PT}$ symmetry.
Correspondingly, $H^{\pm}_{\gamma,\rm surf}$ holds a non-Bloch $\mathcal{CP}$ symmetry, due to the additional imaginary factor $\pm i$.
Therefore, the surface states of $H_{\gamma,\rm 3D}$ can support various NHSE channels activated by $\mathcal{CP}$ symmetry breaking, as discussed in Fig.~\ref{fig3_3Dmain}a to c.

Similarly, the effective surface Hamiltonian of $H_{g,\rm 3D}$ reads
\begin{eqnarray}
	H^{\pm}_{g,\rm surf}&=&\hat{P}_{\pm,{\rm 3D}}H_{g,\rm 3D}\hat{P}_{\pm,{\rm 3D}}=\pm iH_{g,\rm 2D}.\label{eq:prog}	
\end{eqnarray}
Therefore, $H^{\pm}_{g,\rm surf}$ satisfies a $\mathcal{CP}$ symmetry $[\mathcal{CP},H^{\pm}_{g,\rm surf}]_{+}=0$ with $\mathcal{CP}=\sigma_x\mathcal{K}$.
And the surface (hinge) states of $H_{g,\rm 3D}$ displays similar NHSE phenomenon as the bulk( and edge) states of $H_{g,\rm 2D}$,  as shown in Fig.~\ref{fig3_3Dmain}d.

The effective surface Hamiltonian of $H'_{g,\rm 3D}$ is given by
\begin{eqnarray}
	H^{\pm'}_{g,\rm surf}&=&\hat{P}_{\pm,{\rm 3D}}H'_{g,\rm 3D}\hat{P}_{\pm,{\rm 3D}}\nonumber\\
	&=&\pm i[H_{g,\rm 2D}\mp i(t_2'\sin k_y+g)\sigma_0],\label{eq:prog2}	
\end{eqnarray}
where the $\mathcal{CP}$ symmetry is broken and then surface states suffer from NHSE.
However, one branch of boundary states of $H^{\pm'}_{g,\rm surf}$ recovers $\mathcal{CP}$ symmetry and is free from NHSE, analogous with the discussion of $H'_{g,\rm 2D}$ in Eq.~(\ref{eq:H_g}).
In this 3D model, $\mathcal{CP}$ symmetry is recover for distinct branches of boundary states in the $(001)$ and $(00\bar{1})$ surfaces because of the $\mp$ sign in the square bracket of the Eq.\eqref{eq:prog2}. As a result, these NHSE-free hinge states living in these two surfaces are localized in opposite sides along $x$-direction, as shown in Fig.~\ref{fig3_3Dmain}e (green).

Finally, the projectors of $(001)$ and $(00\bar{1})$ surface states for the Hamiltonian $H'_{\rm 3D}$ are $$\hat{P}'_{\pm,{\rm 3D}}=(1\pm \sigma_z\tau_0)/2.$$
Their effective surface Hamiltonian can be obtained as
\begin{eqnarray}
	H^{\pm'}_{3D,\rm surf}&=&\hat{P}'_{\pm,{\rm 3D}}H'_{\rm 3D}\hat{P}
	_{\pm,{\rm 3D}} \nonumber\\
	&=&t_1\cos k_y\tau_0\pm[(u'+v'\cos k_x)\tau_x+v'\sin k_x\tau_y],\nonumber\\ \label{eq:pro3D}	
\end{eqnarray}
which satisfies a $\mathcal{PT}$ symmetry $[\mathcal{PT},H^{\pm'}_{3D,\rm surf}]=0$ with $\mathcal{PT}=\tau_x\mathcal{K}$.
Actually, $H^{\pm'}_{3D,\rm surf}$ is Hermitian and without NHSE.
Correspondingly, the dimensionality of surface and hinge states of $H'_{\rm 3D}$ is unchanged by NHSE, despite that the localizing direction of hinge states (green) are altered by the $z$-NHSE, as shown in Figs.~\ref{fig3_3Dmain}f and \ref{fig3_3Dmain}g.
Meanwhile, the bulk states can show distinct localization with various types of NHSE, as shown in Fig.~\ref{fig3_3Dmain}f and \ref{fig3_3Dmain}g.

\section*{Data availability}
\noindent
Raw numerical data from the plots presented are available from the authors upon reasonable request.

\section*{Code availability}
\noindent
Though not essential to the central conclusions of this work, computer codes for generating our figures are available from the authors upon reasonable request.

\section*{Acknowledgements}
\noindent
This work is
supported by National Natural Science Foundation of China (Grant No. 12104519) and the
Guangdong Project (Grant No. 2021QN02X073). C.H.L acknowledges support from the QEP2.0 Grant from the Singapore National Research Foundation (Grant No. NRF2021QEP2-02-P09) and the Singapore MOE Tier-II Grant (Proposal ID: T2EP50222-0008).
\\

\section*{Author contributions}
\noindent
L.L. initiated this project. 
Z.Lei. conceived the idea of using parity-time symmetry to activate non-Hermitian skin effect along boundaries.
C.H.Lee refined the project and proposed to establish a theoretical framework of selective activated non-Hermitian skin effect.
All authors discussed the theoretical and computational results and contributed to the writing of the manuscript.
\\

\section*{Competing interests}
\noindent
The authors declare no competing interests.
\\

\section*{Additional information}
\noindent
Supplementary information is available for this paper at...
\\

\clearpage

\begin{widetext}

\setcounter{secnumdepth}{2}

	\setcounter{equation}{0} \setcounter{figure}{0} \setcounter{table}{0} %
	\renewcommand{\theequation}{S\arabic{equation}} \renewcommand{\thefigure}{S%
		\arabic{figure}} \renewcommand{\bibnumfmt}[1]{[S#1]}
	\renewcommand\thesection{Supplementary Note \arabic{section}}


\section{Further results of $H_{\gamma,\rm 2D}$}
\subsection{Corner NHSE in the generalized $\mathcal{PT}$ broken regime}\label{appB}
In the main text, we have discussed the 2D Hamiltonian $H_{\gamma,\rm 2D}$ in the generalized $\mathcal{PT}$ symmetry unbroken (Fig.2 of the main text) and intermediate regime (Fig.1 of the main text),
where generalized $\mathcal{PT}$ symmetry is unbroken for either all or some values of $k_y$.
In this section, we discuss another situation where the generalized $\mathcal{PT}$ symmetry is partially broken [$u^2+v^2>(\gamma+2t_2\sin k_y)^2/4>u^2$] for {\it all} values of $k_y$.

\begin{figure}[htb]
	\includegraphics[width=0.4\columnwidth]{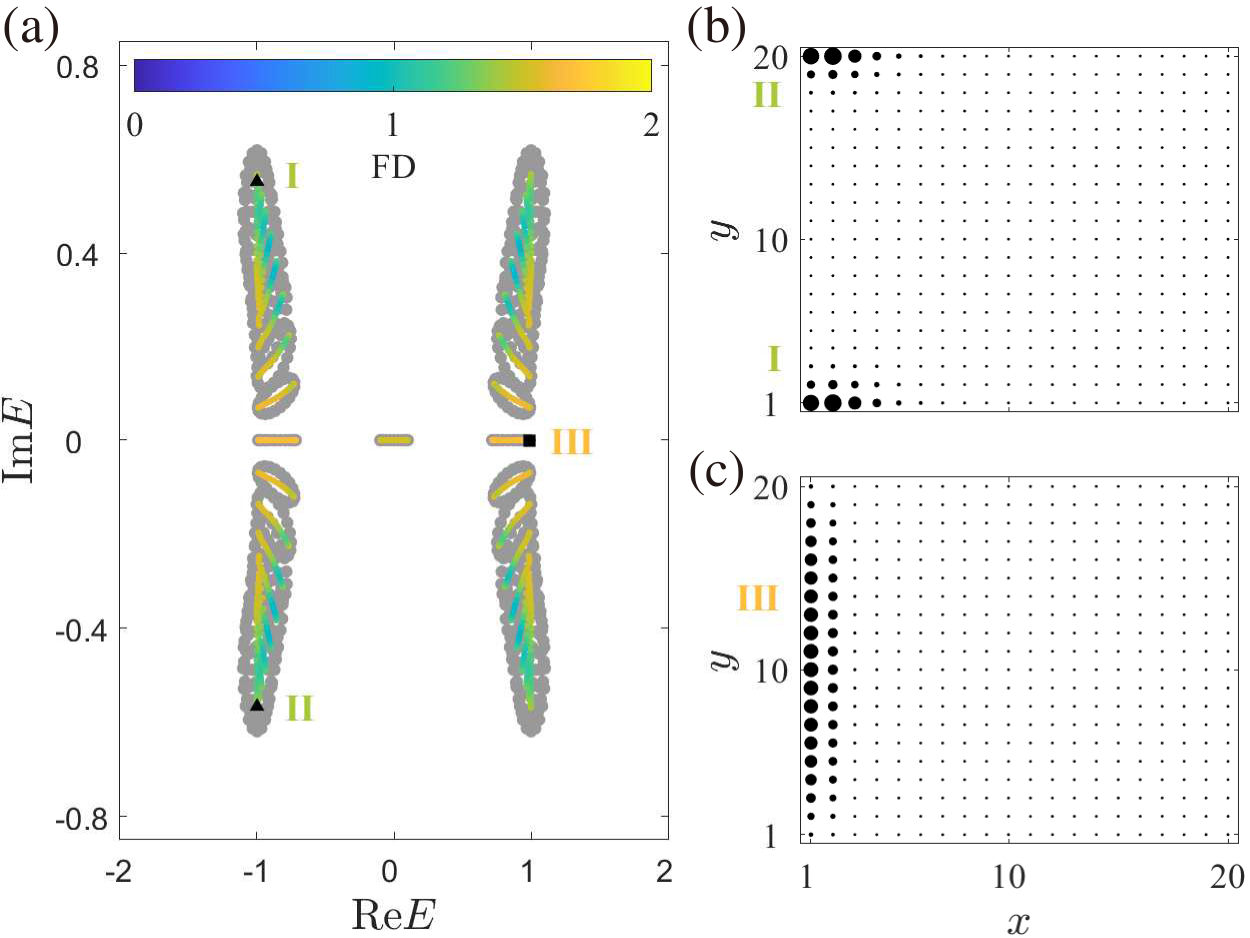}
	\caption{\label{subfig_assi2D_broken} \textbf{The spectrum and distribution of a single eigenstate of $H_{\gamma,\rm 2D}$ with corner NHSE.} (a) $x$-OBC/$y$-PBC (gray) and full OBCs (colored) spectra. Different colors indicate the FD of each eigenstate. (b) and (c) the distributions of eigenstates under full OBCs indexed by triangles and square in (a), respectively. In all panels, the parameters are chosen as $\gamma=1.4$, $t_1=t_2=0.1$, $v=1$, and $u=0.5$, and full OBCs.
	}
\end{figure}

In this regime, almost all bulk states become corner-skin states under the full OBCs, as shown in Fig. \ref{subfig_assi2D_broken}(a) and \ref{subfig_assi2D_broken}(b).
The only exceptions are a $\mathcal{O}(L)$ number of states falling on the real axis of the complex energy plane, which are extended along $y$ direction [Fig. \ref{subfig_assi2D_broken}(c)].
This is because these states correspond to those with $k_x\in\{0,\pi\}$ of the 1D non-Bloch Hamiltonian $\bar{H}_\gamma(k_x)$, which may still hold real eigenenergies even in the generalized $\mathcal{PT}$ broken phase (or, alternatively, the non-Bloch $\mathcal{PT}$ asymmetric regime, See ``Methods'' section of the main text).

\subsection{Evolution process for $H_{\gamma,\rm 2D}$ with mutual exclusively or conventional NHSE}\label{appC}
In this section, we discuss the evolution process governed by Hamiltonian $H_{\gamma,\rm 2D}$.
As in the main text, here the initial state is also chosen to be $|\psi(0)\rangle=\left(|\uparrow,x_0,y_0\rangle+|\downarrow,x_0,y_0\rangle\right)/\sqrt{2}$ with $(x_0,y_0)=(N_x/2,N_y/2)$, and the state at time $t$ is given by $|\psi(t)\rangle=e^{-iH_{\gamma,\rm 2D}t}|\psi(0)\rangle$.

\begin{figure}[htb]
	\includegraphics[width=0.6\columnwidth]{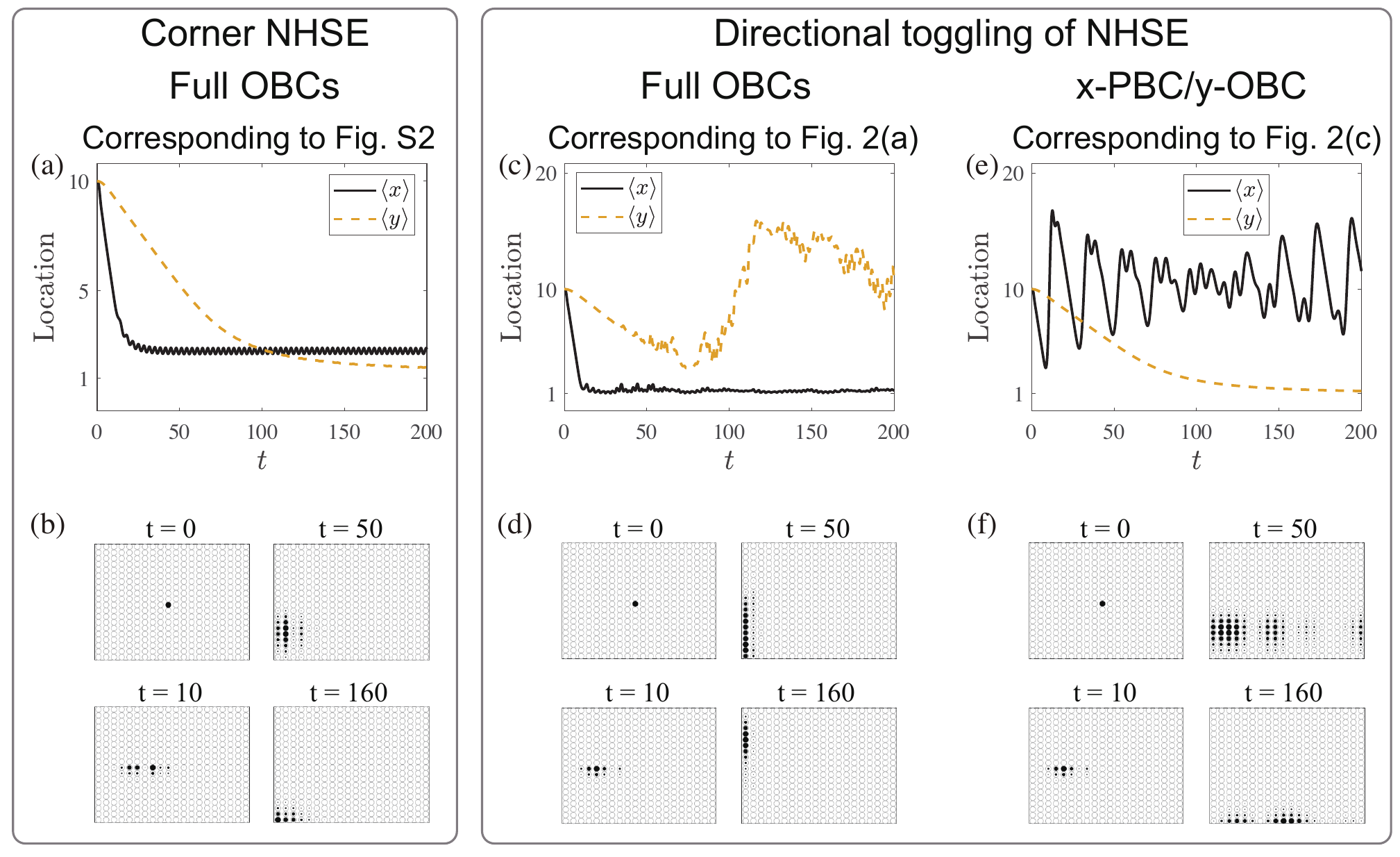}
	\caption{\label{sub_dyna} \textbf{The dynamical process for Hamiltonian $H_{\gamma,\rm 2D}$ with various types of NHSE.} (a), (c), and (e) The mean position of states $\langle x\rangle$ and $\langle y\rangle$ during the evolution process. Here the initial state is prepared in the center of the $2$D lattices. (b), (d), and (f) The normalized distribution of the state $|\psi(t)\rangle$ at different time.
	In (a) and (b), parameters are $\gamma=1.4$, $t_1=t_2=0.1$, $v=1$, $u=0.5$, the same as in Fig. \ref{subfig_assi2D_broken} where corner NHSE occurs.
	(c) to (f) have the same parameters except for $u=0.9$, the same as in Fig. 2 of the main text corresponding to generalized $\mathcal{PT}$ unbroken phase.
	The boundary conditions are chosen to be full OBCs in (a) to (d), and $x$-PBC/$y$-OBC in (e) and (f). In all panels the system's size is $N_x=N_y=20$.
	}
\end{figure}

We first discuss the regime $u<\gamma/2-t_2$ where
the generalized PT symmetry is broken and almost all bulk state suffer from corner NHSE.
Specifically, bulk states with positive imaginary energy (which shall dominate the dynamics) locate at the bottom left corner, as discussed in Sec~\ref{appB} and shown in Fig. \ref{subfig_assi2D_broken}(a) and \ref{subfig_assi2D_broken}(b).
As a result, $\langle x\rangle$ and $\langle y\rangle$ decreases rapidly to $\sim1$, as shown in Fig. \ref{sub_dyna}(a).
Figure \ref{sub_dyna}(b) further demonstrates the distribution of $|\psi(t)\rangle$ at different values of $t$.

In contrast, bulk states of this model locates at left edge under full OBCs when the generalized $\mathcal{PT}$ symmetry is unbroken, as shown in Fig.2(a) in the main text.
Correspondingly, $\langle x\rangle$ decreases rapidly while $\langle y\rangle$ does not show a clear increasing or decreasing tendency, as shown in Fig. \ref{sub_dyna}(c).
Consistently, a state prepared deep in the bulk will evolve to the left boundary and remain there in a long-time dynamics, as shown in Fig. \ref{sub_dyna}(d).

Finally, we demonstrate the dynamics for the same parameters as in Fig. \ref{sub_dyna}(c) and (d), but with $x$-PBC/$y$-OBC instead of full OBCs.
As discussed in the main text, a $y$-NHSE shall emerge in this scenario.
Specifically, bulk states with positive imaginary parts distribute along the bottom edge, as shown in Fig.2(c) of the main text.
Therefore, after replacing $x$-OBC with $x$-PBC, $\langle y\rangle$ decreases rapidly while $\langle x\rangle$ oscillates around $N_x/2$, as shown in Fig. \ref{sub_dyna}(e).
More explicitly, evolution of a state behaves the same under the full OBCs and $x$-PBC/$y$-OBC before it reaches a boundary of the system, as shown in Fig. \ref{sub_dyna}(d) and Fig. \ref{sub_dyna}(f) with $t=10$.
Afterward, different boundary conditions lead to distinct dynamical behaviors, and $|\psi(t)\rangle$ remains in the bottom boundary in a long-time dynamics for the system under $x$-PBC/$y$-OBC [Fig. \ref{sub_dyna}(f)].

\section{Geometry-independence of the intrinsic hybrid skin-topological effects}\label{appD}
In this section, we discuss the intrinsic hybrid skin-topological effect of $H_{g,\rm 2D}$ in Eq.(5) of the main text with different geometries.
As shown in Fig. 3(a) of the main text, the bulk spectrum of $H_{g,\rm 2D}$ is given by 1D lines and covers zero area in the complex plane.
As discussed in Ref.~\cite{Zhang2022}, this property reveals that the absence of bulk NHSE is independent from the system's geometry, hence we denote it as the intrinsic hybrid skin-topological effect.
To verify this, we demonstrate the complex spectrum and spatial distribution of eigenstates of Hamiltonian $H_{g,\rm 2D}$ under triangle geometry in Fig. \ref{supfig_geo}.
As seen in Fig. \ref{supfig_geo}(a), the bulk spectrum (black) is still consistent with the $x$-OBC/$y$-PBC spectrum (gray).
And as further demonstrated in Fig. \ref{supfig_geo}(b),
bulk states are extended in the triangle lattice and are free from NHSE, while edge states suffer from NHSE and are located at corners.

\begin{figure}[htb]
	\includegraphics[width=0.5\columnwidth]{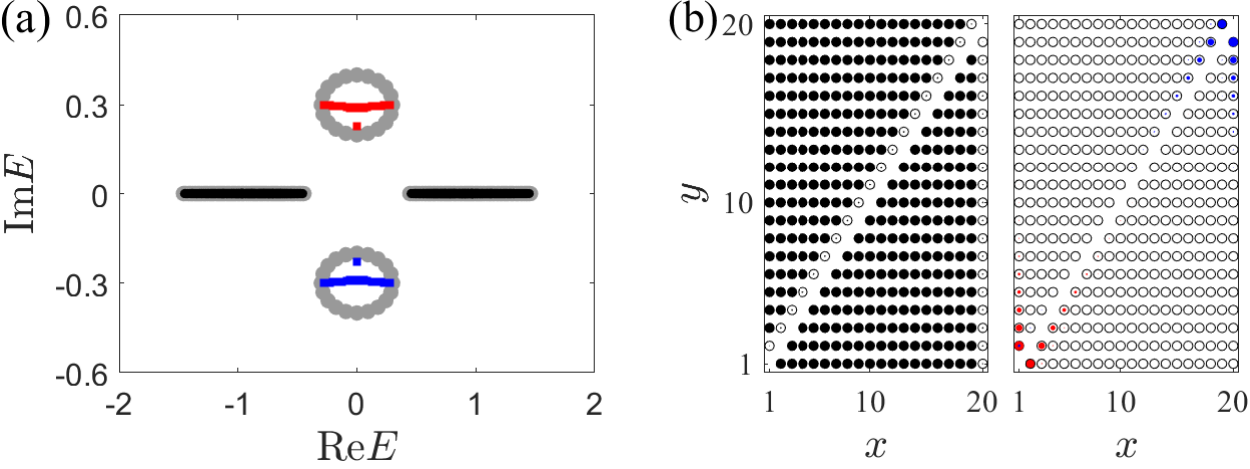}
	\caption{\label{supfig_geo} \textbf{Geometry-independence of intrinsic hybrid skin-topological effect of $H_{g,\rm 2D}$.} (a) Energy spectrum for Hamiltonian $H_{g,\rm 2D}$ in Eq.(5) of the main text under triangle geometry. Black circles correspond to bulk states and red (blue) squares correspond to boundary states with positive (negative) imaginary energies.
		Gray circles show $x$-OBC/$y$-PBC spectra for the same Hamiltonian.
		(b) Spatial distribution of states under triangle geometry, marked by the same colors as in (a). Parameters are $u=0.2$, $g=0.3$, $t_1=0.3$ and $t'_2=0.1$ for all panels.}
\end{figure}

\begin{figure}[htb]
	\includegraphics[width=0.5\columnwidth]{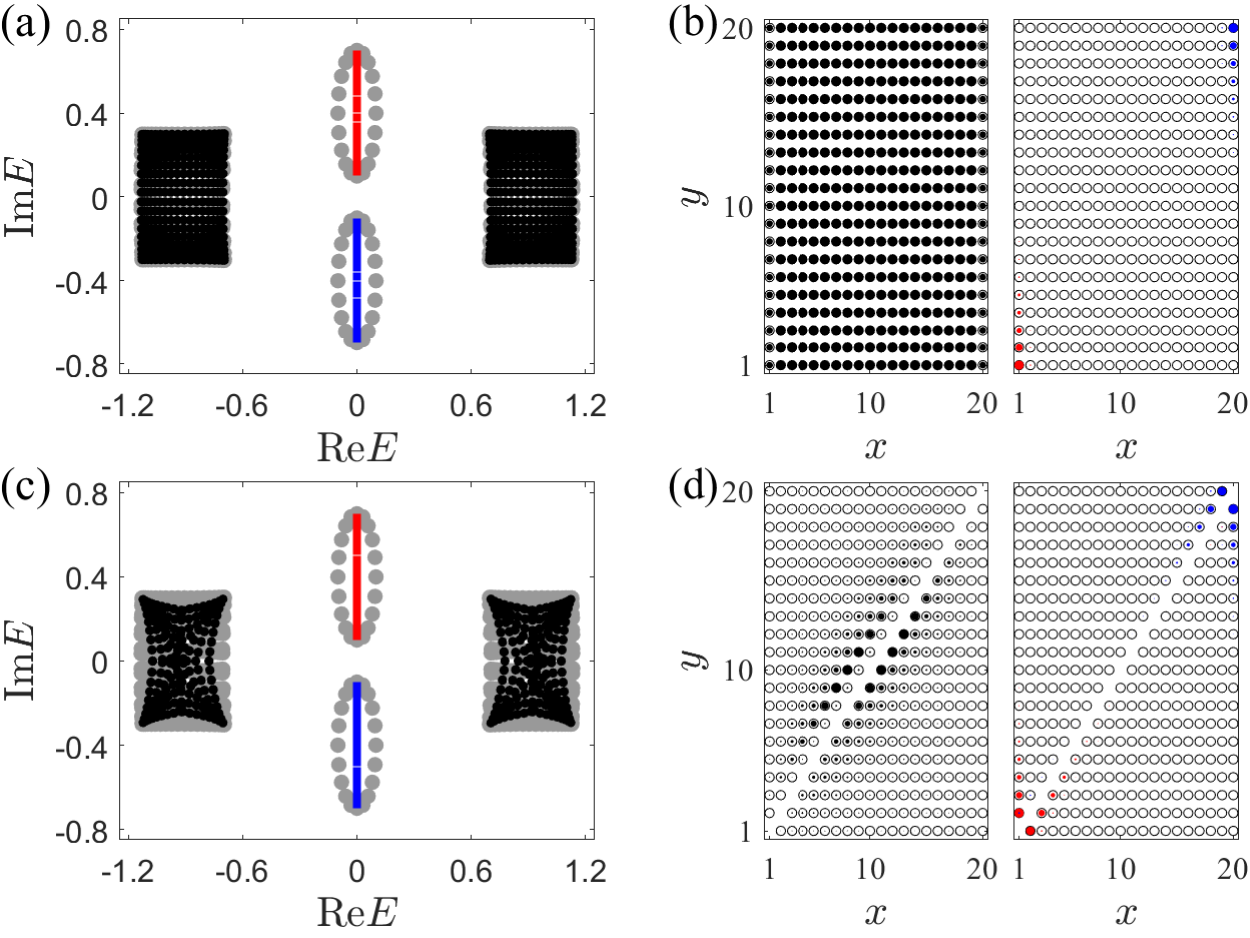}
	\caption{\label{supfig_geocom} \textbf{Geometry-dependence of conventional hybrid skin-topological effect of $H^{(\rm geo)}_{g,\rm 2D}$.} (a) Energy spectrum of Hamiltonian $H^{(\rm geo)}_{g,\rm 2D}$ in Eq.~\eqref{2D_H_geode} under square geometry.
		Black circles corresponds to bulk states and red (blue) squares corresponds to boundary states with positive (negative) imaginary energies.
		Gray circles shows $x$-OBC/$y$-PBC spectrum for the same Hamiltonian.
		(b) Spatial distribution for states under square geometry, marked by the same colors as in (a). (c) and (d) the same as (a) and (b), but under triangle geometry. Parameters are $u=0.2$, $g=0.3$, $t_1=0.3$ and $t'_2=0.1$ for all panels.}
\end{figure}
 \end{widetext}

In contrast, we can construct another Hamiltonian as followed,
\begin{eqnarray}
	H^{(\rm geo)}_{g,\rm 2D}=H_{g}+t'_2\cos k_y\sigma_z+it_1\sin k_y\sigma_0,
	\label{2D_H_geode}
\end{eqnarray}
which is similar to Hamiltonian $H_{g,\rm 2D}$ but with an extra $\pi$ phase to the hopping toward positive $y$ direction.
This Hamiltonian also supports hybrid skin-topological effect under square geometry, as demonstrated in Fig. \ref{supfig_geocom} (a) and \ref{supfig_geocom} (b).
However, its bulk spectrum covers a non-vanishing area in the complex plane, suggesting the emergence of bulk NHSE on certain geometries.
Specifically, we find that the bulk states suffer from NHSE under triangle geometry, localizing along the hypotenuse of the triangle, as demonstrated in Fig. \ref{supfig_geocom} (c) and (d).



\end{document}